\documentclass[twocolumn,showpacs,preprintnumbers]{revtex4}
\usepackage{amsfonts}
\usepackage{amsmath}
\usepackage{graphicx}
\usepackage{dcolumn}
\usepackage{bm}

\input{tcilatex}

\begin{document}

\preprint{}
\title{Two components of dark matter in the DAMA data}
\author{Yukio Tomozawa}
\affiliation{Michigan Center for Theoretical Physics, Randall Laboratory of Physics,
University of Michigan, Ann Arbor, MI. 48109-1040}
\date{\today }

\begin{abstract}
It is shown that the DAMA data indicate two dark matter components, one that
circulates around the galactic center (GC) and another that is emitted from
the GC. From the location of the maximum yearly variation, one can compute
the ratio of the two components.
\end{abstract}

\pacs{04.70.-s, 95.85.Pw, 95.85.Ry, 98.54.Cm}
\maketitle

\section{Introduction}

The yearly variation observed in the DAMA experiment\cite{dama},\cite%
{dmsearch} extends for 12 years and the location of the maximum has been
established with 8.9$\sigma $. Since the time at the maximum does not
coincide with that when the Earth orbit approaches the GC or is tangential
to a circular orbit around the GC, it is natural to assume that dark matter
has two components, one that circulates around the GC and another that is
emitted from the GC. This article computes the ratio of the two cmponents,
and explores its significance. A possibility of the Migdal effect being a
source that differentiates the results of the DAMA data and the CDMS data is
discussed.

\section{Two dark matter particle (DMP) components}

If the GC is on the 18h of the RA coordinate, the Earth is moving towards it
on March 22nd (81 days from January 1st) and moves tangentially to a
circular orbit around the GC on June 22nd (173 days) and December 22. The
real coordinates of the GC are\cite{allen} 
\begin{eqnarray}
RA &=&17^{h}45^{m}37^{s}.1991=266^{\circ }.40499625 \\
DEC &=&-28^{\circ }56^{\prime }10".221=-28^{\circ }.93617242\text{ \ \ \ }%
(J2000),
\end{eqnarray}%
then the time when the Earth is moving towards the GC is March 18.35 or%
\begin{equation}
t_{01}=77.35\text{ \ \ \ \ \ }in\text{ }days  \label{eq1}
\end{equation}%
and the time when the Earth is closest to the GC is%
\begin{equation}
t_{02}=168.66\text{ \ \ \ \ }in\text{ }days.  \label{eq2}
\end{equation}%
Since neither of these dates coincides with the observed time of maximum
yearly variation\cite{dama},%
\begin{equation}
t_{0}=145\pm 7,\text{ \ \ \ \ }(8.9\sigma ),  \label{eq3}
\end{equation}%
one has to assume that there are two dark matter components. Let us denote
the yearly intensity variation of the component coming from the GC by $A_{1}$
and that of the component along a circular orbit by $A_{2}$. The total DMP
intensity variation that yields $A$ is given by%
\begin{equation}
A\cos (\omega (t-t_{0}))=A_{1}\cos (\omega (t-t_{01}))+A_{2}\cos (\omega
(t-t_{02})),
\end{equation}%
which gives%
\begin{equation}
A\cos (\omega t_{0})=A_{1}\cos (\omega t_{01})+A_{2}\cos (\omega t_{02})
\label{eq4}
\end{equation}%
and%
\begin{equation}
A\sin (\omega t_{0})=A_{1}\sin (\omega t_{01})+A_{2}\sin (\omega t_{02}),
\label{eq5}
\end{equation}%
where%
\begin{equation}
\omega =\frac{2\pi }{365.2422}\text{ \ \ \ \ }(in\text{ }days^{-1}).
\end{equation}

From Eq. (\ref{eq4}) and Eq. (\ref{eq5}), one gets%
\begin{eqnarray}
A^{2} &=&A_{1}^{2}+A_{2}^{2}+2A_{1}A_{2}\cos (\omega (t_{02}-t_{01})) \\
&=&A_{1}^{2}+A_{2}^{2},
\end{eqnarray}%
since%
\begin{equation}
\omega (t_{02}-t_{01})=\frac{\pi }{2},
\end{equation}%
and%
\begin{equation}
\tan (\omega t_{0})=\frac{A_{1}\sin (\omega t_{01})+A_{2}\sin (\omega t_{02})%
}{A_{1}\cos (\omega t_{01})+A_{2}\cos (\omega t_{02})}.  \label{eq6}
\end{equation}%
Solving Eq. (\ref{eq6}) for $A_{2}/A_{1}$, one obtains%
\begin{equation}
\frac{A_{2}}{A_{1}}=\frac{\sin (\omega (t_{0}-t_{01}))}{\sin (\omega
(t_{02}-t_{0}))}.
\end{equation}%
From Eq. (\ref{eq1}-\ref{eq3}), one gets%
\begin{equation}
\frac{A_{2}}{A_{1}}=2.35\pm 1.04.  \label{eq7}
\end{equation}

The DMP component that circulates on a circular orbit around the GC is
moving with the solar system and it creates a yearly variation due to a
variation in the radial DMP distribution, as is described in the next
section. This gives the $A_{2}$ component. In March, when the Earth is
approaching the GC, it encounters DMP emitted from black holes at the GC
with higher relative velocity than in September, when the earth is moving
away from the GC, with smaller relative velocity. This gives the $A_{1}$
component. The details of the both components will be described in the
following sections.

\section{DMP circulating the GC}

The simplest assumption that gives the flat velocity distribution observed
in most spiral galaxies is a spherically symmetric distribution of DMP. If
one assumes that DMP are orbiting in all possible directions, no yearly
variation is produced, since a summer-winter difference is cancelled by two
DMP orbitting in the opposite directions. A yearly variation is produced by
a radial distribution of DMP. If the DMP distribution is a decreasing
function of distance from the GC, then the Earth encounters more DMP flux in
the summer due to the fact that Earth is closer to the GC in the summer.
This is consistent with the DAMA data in the previous section.

If the orbital motion of DMP is constrained so that the angular momentum
component relative to that of the solar system or the visible spiral
galactic plane, then the component of the orbital motion in the galactic
plane contributes to the yearly varition, since the component vertical to
the galactic plane cancels between orbits with opposite vertical components.
Assuming that the average orbit is circular, its component in the galactic
plane has a velocity less than that of the solar system. Then the maximum of
yearly variation occurs in the winter, since the orbital motion in the
vertical plane is opposite to that of the solar system relative to the
latter, in contradiction to the computation in the previous section. In
order to get a solution with a maximum in the summer, one has to have a
contribution from a decreasing radial distribution compensating the effect
of orbital motion.

In both cases, the effect of the radial distribution has to dominate the
yearly variation.

\section{Emission of DMP from GC black holes}

The existence of DMP in orbital motion around the GC is quite natural. It is
also natural to assume that its distribution is a decreasing function of
distance from the GC at the location of the solar system, in order to get a
flat velocity distribution around the GC. However, the existence of DMP
emitted from the GC requires an explanation. That is related to a recent
discovery by the Pierre Auger Project concerning the correlation between
high energy cosmic rays and the location of AGN (Active Galactic Nuclei)\cite%
{auger} and a model of the author which predicted the Auger data since 1985.

The Pierre Auger Project data suggest that high energy cosmic rays are
emitted from AGN, massive black holes. In a series of articles\cite{cr1}-%
\cite{cr9}, the author has presented a model for the emission of high energy
particles from AGN. The following is a summary of the model.

1) Quantum effects on gravity yield repulsive forces at short distances\cite%
{cr1},\cite{cr3}.

2) The collapse of black holes results in explosive bounce back motion with
the emission of high energy particles.

3) Consideration of the Penrose diagram eliminates the horizon problem for
black holes\cite{cr4}. Black holes are not black anymore.

4) The knee energy for high energy cosmic rays can be understood as a split
between a radiation-dominated region and a matter-dominated region, not
unlike that in the expansion of the universe. (See page 10 of the lecture
notes\cite{cr1}-\cite{cr3}.)

5) Neutrinos and gamma rays as well as cosmic rays should have the same
spectral index for each AGN. They should show a knee energy phenomenon, a
break in the energy spectral index, similar to that for the cosmic ray
energy spectrum.

6) The recent announcement by Hawking rescinding an earlier claim about the
information paradox\cite{hawking} is consistent with this model.

Further discussion of the knee energy in the model predicts the existence of
a new mass scale in the knee energy range, in order to have the knee energy
phenomenon in cosmic rays \cite{crnew}. The following are additional
features of the model.

7) If the proposed new particle with mass in the knee energy range (0.1 PeV$%
\sim $2 PeV) is stable and weakly interacting with ordinary particles, then
it becomes a candidate for a DMP. It does not necessarily have to be a
supersymmetric particle. That is an open question. However, if it is
supersymmetric, then it is easy to make a model for a weakly interacting DMP%
\cite{pevss}. The only requirement is that such particles must be present in
AGN or black holes so that the the knee energy is observed when cosmic rays
are emitted from AGN. A suggested name for the particle is sion (xion)\cite%
{crnew}, using the Chinese/Japanese word for knee, si (xi).

8) If the particle is weakly interacting, then it does not obey the GZK
cutoff\cite{gzk}, since its interaction with photons in cosmic backgroud
radiation is weak, as was pointed out earlier. This is a possibe resolution
of the GZK puzzle\cite{agasa},\cite{puzzle},\cite{crnew2}.

In summary, this model predicted the Pierre Auger Project data. Moreover it
suggests the existence of a new particle in the PeV mass range, in order to
explain the knee energy phenomenon in the cosmic ray spectrum. The author
has suggested that this new particle is a candidate for DMP\cite{crnew3}.

\section{AGN and GC emits DMP}

We assume that the incident particles above the GZK cutoff observed by the
Akeno-AGASA detector are weakly interacting particles at the PeV mass scale,
which are required to exist in order to explain the cosmic ray knee energy
in the model. In a model where the acceleration takes place by gravity such
as that proposed by the author, there is no difficulty in accelerating a
weakly interacting and neutral DMP by gravity.

Then AGASA and DAMA data present a consistent picture for DMP emission from
black holes. The only difference is the AGASA data captures the high energy
end of the DMP spectrum, while the DAMA data captures the lower end: In the
latter, the cross section increases linearly with energy and the energy
spectrum of the particle distribution decreases with energy, $E^{-2.5}%
\symbol{126}E^{-3}$, so that the lower end of the spectrum contributes to
the DAMA data.

The discussion in the previous section and this section provides a basis for
the DMP component emitted from the GC that is implied by the DAMA data. Due
to the nature of weak interactions, where cross sections are an increasing
function of energy, March has a higher probability of encountering with the
dark matter than September.

\section{The Migdal effect and the DAMA data}

An important question is why other experiments such as CDMS have not
observed a yearly variation of DMP. It was suggested that enhancement by the
Migdal effect might be a key factor\cite{migdal}, \cite{migdal2}. In ref. 
\cite{migdal}, the Migdal effect is computed for a NaI detector and it is
shown that there is a large ionization effect for the inner electron. The
Migdal effect is known to give a smaller effect for targets with large
atomic number Z, since the Coulomb binding is proportional to (Ze)$^{2}$ and
therefore the Migdal effect is proportional\cite{migdal2} to $1/Z^{2}$. The
Migdal effect on the Ge target of CDMS is 1/10 of that of Na target of DAMA.
That might explain the different results for the two detectors.

\begin{acknowledgments}
\bigskip The author would like to thank David N. Williams for reading the
manuscript.
\end{acknowledgments}

\bigskip

\bigskip


\bigskip

\begin{thebibliography}{99}
\bibitem{dama} Bernabei, R. et al., First results from DAMA/LIBRA and the
combined results with DAMA/NaI, arXiv:0804.2741.

\bibitem{dmsearch} DM 2008, Dark Matter and Dark Energy in the Universe,
Feb. 20-22 (2008), California, http://ppd.fnal.gov/experiments/cdms/

\bibitem{allen} Cox, N. Arthur, Editor, Allen's Astrophysical Quantities, 4
th edition, (AIP Press, Springer-Verlag, N.Y., 2000) p. 575.

\bibitem{auger} The Pierre Auger Collaboration, Science 318, 938 (2007);
Correlation of the Highest-energy Cosmic Rays with the Positions of Nearby
Active Galactic Nuclei, arXiv: 0712.2843 (2007).

\bibitem{cr1} Tomozawa, Y., Magnetic Monopoles, Cosmic Rays and Quantum
Gravity, in the Proc. of 1985 INS International Symposium on Composite
Models of Quarks and Leptons (Tokyo, edit. Terazawa, H. and Yasue, M.,
1985), pp. 386.

\bibitem{cr2} Tomozawa, Y., The Origins of Cosmic Rays and Quantum Effects
of Gravity, in Quantum Field Theory (ed. Mancini, F., Ersever Science
Publishers B. V., 1986) pp. 241. This book is the Proceedings of the
International Symposium in honor of Hiroomi Umezawa held in Positano,
Salerno, Italy, June 5-7, 1985.

\bibitem{cr3} Tomozawa, Y., Cosmic Rays, Quantum Effects on Gravity and
Gravitational Collapse, Lectures given at the Second Workshop on Fundamental
Physics, University of Puerto Rico, Humacao, March 24-28, 1986. This lecture
note can be retrieved from KEK Kiss NO 200035789 at
http://www-lib.kek.jp/KISS/kiss\_prepri.html

\bibitem{cr4} Tomozawa, Y., Gravitational Waves, Supernova and Quantum
Gravity, in Symmetry in Nature, (Scuola Normale Superiore, Pisa, 1989) pp.
779, Section 2 and 3.

\bibitem{cr5} Tomozawa, Y., Exact Solution of the Quantum Einstein Equation
and the Nature of Singularity, in the Proc. 5th Marcel Grossman Meeting on
General Relativity (ed. D. Blair et al., Perth, Australia, 1988) pp. 527.

\bibitem{cr6} Tomozawa, Y., Black Hole Oscillation, in the Proc. 5th Marcel
Grossman Meeting on General Relativity (ed. D. Blair et al., Perth,
Australia, 1988) pp. 629.

\bibitem{cr7} Majumdar, A. and Tomozawa, Y., Progr. Theoret. Phys. (Kyoto)
82, 555 (1989).

\bibitem{cr8} Majumdar, A. and Tomozawa, Y., Nuovo Cimento 197B, 923 (1992).

\bibitem{cr9} Tomozawa, Y., Astron. Astrophys. Suppl. Ser. 97, 117 (1993).

\bibitem{hawking} Hawking, S.,
http://www.newscientist.com/article/dn6151.html.

\bibitem{crnew} Tomozawa, Y., High Energy Cosmic Rays, Gamma Rays and
Neutrinos from AGN, arXiv: 0802.0301 (2008).

\bibitem{pevss} Wells, J. D., Phys. Rev D71, 015013 (2005)

\bibitem{gzk} .Greisen, K., Phys. Rev. Lett. 16, 748 (1966); Zatsepin, G. T.
and Kuzmin, V. A., Pisma Z. Experim. Theor. Phys. 4, 114 (1966).

\bibitem{agasa} Shinozaki, K. et al., Nucl. Phys. B (Proc. Suppl.) 136, 18
(2004); Nagano, M. and Watson, A. A., Rev. Mod. Phys. 72, 689 (2000).

\bibitem{puzzle} Berezinsky, V., Astroparticle Physics: Puzzles and
Discoveries, arXiv:0801.3028 (2008).

\bibitem{crnew2} Tomozawa, Y., High Energy Cosmic Rays from AGN and the GZK\
Cutoff, arXiv: 0802.2927 (2008).

\bibitem{crnew3} Tomozawa, Y., Search for a dark matter particle in high
energy cosmic rays, arXiv: 0804.1499 (2008).

\bibitem{migdal} Bernabei, R. et al., Int. J. of Mod. Phys. A 22, 3155 (2007)

\bibitem{migdal2} Migdal, A. B., J. Phys. USSR 4, 449 (1941); Migdal, A. B.,
Qualitative Methods in Quantum Mechanics (Benjamin, Reading, Massachusetts,
1977), p. 108; Landau, L. D. and Lifshits, E. M., Quantum Mechanics,
Non-Relativistic Theory, 3rd ed. (Pergamon Press, 1977), \ p. 149.
\end{thebibliography}
\end{document}